# Solar Modulation of Cosmic Rays during the Declining and Minimum Phases of Solar Cycle 23: Comparison with Past Three Solar Cycles

O.P.M. Aslam · Badruddin
Department of Physics, Aligarh Muslim University, Aligarh-202002, India.
e-mail: badr.physamu@gmail.com

**Abstract** We study solar modulation of galactic cosmic rays (GCRs) during the deep solar minimum, including the declining phase, of solar cycle 23 and compare the results of this unusual period with the results obtained during similar phases of the previous solar cycles 20, 21, and 22. These periods consist of two epochs each of negative and positive polarities of the heliospheric magnetic field from the north polar region of the Sun. In addition to cosmic ray data, we utilize simultaneous solar and interplanetary plasma/field data including the tilt angle of the heliospheric current sheet. We study the relation between simultaneous variations in cosmic ray intensity and solar/interplanetary parameters during the declining and the minimum phases of cycle 23. We compare these relations with those obtained for the same phases in the three previous solar cycles. We observe certain peculiar features in cosmic ray modulation during the minimum of solar cycle 23 including the record high GCR intensity. We find, during this unusual minimum, that the correlation of GCR intensity is poor with sunspot number ($R = -0.41$), better with interplanetary magnetic field ($R = -0.66$), still better with solar wind velocity ($R = -0.80$) and much better with the tilt angle of the heliospheric current sheet ($R = -0.92$). In our view, it is not the diffusion or the drift alone, but the solar wind convection is the most likely additional effect responsible for the record high GCR intensity observed during the deep minimum of solar cycle 23.

**Keywords** Cosmic ray modulation · Heliospheric current sheet · Solar minimum · Solar wind

## 1. Introduction

Study of cosmic-ray intensity variations at different time scales and under various conditions of solar magnetic activity is useful to understand the physics of interactions between charged particles and the magnetic field or background plasma, dynamics of the heliospheric structures responsible for these variations, and to understand the physical mechanism(s) playing an important role in the modulation of galactic cosmic rays (GCRs; see, *e.g.,* Venkatesan and Badruddin, 1990; Cane, 1993; Potgieter, Burger, and Ferreira, 2001; Kudela *et al.,* 2000; Kudela, 2009; Richardson, 2004; Heber, 2011). Although the solar modulation of galactic cosmic rays has been studied for several decades, it is still a subject of intense research to assess the continuously changing behaviour of the Sun and its influence on cosmic rays.

It is known that galactic cosmic rays are subjected to heliospheric modulation, under the influence of solar output and its variations. This modulation of cosmic ray intensity, associated with the 11-year solar activity cycle and anti-correlated with solar activity, was first studied by Forbush (1958) and by many subsequent researchers (*e.g.,* Burlaga *et al.,* 1985; Storini *et al.,* 1995; Ahluwalia and Wilson, 1996; Mavromichalaki, Belehaki, and Rafios, 1998; Cliver and Ling, 2001; Kane, 2003; Sabbah and Rybansky, 2006; Badruddin, Singh, and Singh, 2007; Singh, Singh, and Badruddin, 2008; and many others). The solar magnetic field reverses at each solar activity maximum resulting in the 22-year cycle as well (Jokipii and Thomas, 1981; Potgieter and Moraal, 1985). The field orientation (polarity) is defined positive when the field is outward from the Sun in the northern hemisphere (*e.g.,* during the 1970's and 1990's) and negative when the field is outward in the southern hemisphere (*e.g.,* during the 1960's and 1980's). A positive polarity field is denoted by the *A*>0 epoch and a negative field by the *A*<0 epoch (Duldig, 2001; Singh and Badruddin, 2006).

Cosmic ray particles in the solar wind plasma flow and the magnetic field of the heliosphere are subjected to four distinct transport effects. (1) An outward convection caused by radially-directed solar wind velocity (convection). (2) The magnetic field varies systematically over large scales, so there are, in addition, curvature and gradient drifts (drifts). (3) Depending on the sign of the divergence of solar wind velocity, there will be energy change (adiabatic energy change). (4) The spatial diffusion is caused by the scattering by random magnetic irregularities (diffusion). (See, *e.g.,* Heber (2011) and references therein). The resulting transport is a superposition of these coherent and random effects.

It is generally accepted (McDonald, Nand Lal, and McGuire, 1993; Potgieter, 1994) that all the above processes are important, but their relative importance varies throughout the solar cycle. In the period near solar minimum, drifts may play an important role in the transport of cosmic rays through the heliosphere. In a complex magnetic structure characteristic of solar maximum, it is likely that drifts play only a small role and that the modulation is determined by large-scale disturbances in the solar wind (McKibben *et al.,* 1995).

The diffusion and convection components of cosmic-ray transport equation are independent of the solar polarity and will only vary with the solar activity cycle. Conversely, the drift components will have opposite effect in each activity cycle following the field reversal around each solar maximum. Positively-charged cosmic-ray particles would essentially enter the heliosphere along the helio-equator and exit via the poles in the *A*<0 polarity state. In the *A*>0 polarity state the flow would be reversed, with particles entering over the poles and exiting along the equator.

Thus, it may be expected that the response of GCR intensity to the changes in solar activity during the *A*<0 epochs is different from that in the *A*>0 solar polarity epochs, which has been observed in a long-term record of cosmic ray intensity. Differences in the response of solar activity to GCR intensity variations (Van Allen, 2000; Singh, Singh, and Badruddin, 2008) and to the tilt angle of the HCS (Badruddin, Singh, and Singh, 2007) during the increasing phases of solar cycles 21, 23 (*A*>0) and 20, 22 (*A*<0) have been noted. Thus it will be interesting to study the response of solar and interplanetary parameters to GCR intensity during different polarity epochs (*A*<0 and *A*>0) when solar activity is decreasing *i.e.,* in the declining (including minimum) phases of different solar cycles.

The recent solar minimum of cycle 23 has been unusually long and deep. In comparison with the previous three minima, this solar minimum has the smallest sunspot number, the lowest, least dense, and coolest solar wind, weaker solar wind dynamic pressure and the weakest solar and interplanetary magnetic field. However, in contrast to the previous minimum, there are more stream interaction regions (SIRs) and more shocks and coronal mass ejections (CMEs). But the ICMEs, which are interplanetary manifestations of CMEs, the SIRs and shocks during this minimum are generally weaker than during the previous minimum (Jian, Russell, and Luhmann, 2011). The average CME mass is also smaller during the recent minimum than the previous one (Vourdilas *et al.,* 2010). However, the heliospheric current sheet (HCS) during this minimum was less flat (more warped) than the previous two minima. All these peculiar solar and heliospheric conditions make this recent minimum an interesting period to study, in particular, from the point of view of cosmic ray modulation.

In this work we have analyzed solar and interplanetary plasma and magnetic field data (http://omniweb. gsfc.nasa.gov) and cosmic ray intensity as observed by a neutron monitor (http://cosmicrays.oulu.fi/) at Oulu for four solar cycles 20, 21, 22, and 23.

## 2. Results and Discussion

### 2.1. Time Variation of Solar and Interplanetary parameters, and GCR Intensity during the Declining and Minimum Phases

The Sun is the dominant variable force, which controls the structure of the heliosphere and the modulation of cosmic rays through the level of solar activity, the tilt angle of the heliospheric current sheet, the velocity of the solar wind, and the strength and turbulence of the interplanetary magnetic field (McDonald, Webber, and Reames, 2010). It will be interesting to compare the nature and magnitude of variabilities in solar activity (*e.g.,* sunspot number), solar wind velocity, interplanetary magnetic field strength and its fluctuations, and the tilt of the heliospheric current sheet during similar phases (*e.g.,* decreasing/minimum) of different solar cycles. To understand the cosmic ray modulation process, it will be more interesting to compare these variabilities with the variability in cosmic ray intensity during the same phase of different solar cycles.

The sunspot number (SSN), the oldest directly observed solar activity on the photosphere, is a very useful indicator of solar activity. The 10.7cm solar radio flux is an indicator of activity in the upper chromosphere and lower corona. Both these solar activity indices are, in general, inversely correlated with the cosmic ray intensity over a long time scale. Out of three GCR modulation processes (convection, diffusion, and drifts), the solar wind velocity ($V$) is related to convection, diffusion depends on the interplanetary field strength ($B$) and its fluctuations $\sigma_B$ (which is considered as a measure of fluctuations in the interplanetary magnetic field), and the tilt of the heliospheric current sheet ( ) is connected to the gradient and curvature drift effects in the large scale magnetic fields. Thus, we have utilized these solar (SSN, 10.7cm flux) and interplanetary ($V$, $B$, $_B$) parameters, including the tilt angle ( ) together with the neutron monitor count rate at Oulu (cut-off rigidity $R_c$ = 0.81 GV)

The level of solar activity is traditionally represented by sunspot numbers. A comparison of sunspot numbers during the declining and minimum phases of four solar cycles 20, 21, 22, and 23 is shown in Figure 1 (a). To compare the four minima, in particular, time zero in this figure corresponds to the end of minimum of a solar activity cycle (after which the solar activity starts increasing for the next solar cycle). The negative time (in solar rotations) means the number of solar rotations before the end of minimum (zero time) in each solar cycle. As compared to the previous three solar cycles, the decline of solar activity in cycle 23 toward minimum is the longest and the weakest one. A similar variability in another solar parameter (10.7cm solar flux) is evident, as shown in Figure 1 (b). The 10.7cm solar flux is given in solar flux units (sfu) (1 sfu = $10^{-22} Wm^{-2} Hz^{-1}$).

The variations in the solar wind speed in the same periods are plotted in Figure 1 (c). Although the variations in the solar wind speed do not strictly follow the sunspot number variations, the speed is slowest during the latest minimum of cycle 23, in comparison to the previous three minima. When similar phases of four solar cycles are compared (see Figure 1 (d)), the interplanetary magnetic field (IMF) shows a declining trend at least during cycles 21, 22, and 23, somewhat similar to the sunspot number. However, such a trend is not clearly seen in the IMF during cycle 20. Further, the magnetic field is weakest during the cycle 23 minimum. These parameters, and many more, are summarized during the deep minimum of solar cycle 23 by Jian, Russell, and Luhmann (2011). However, in contrast to the sunspot number, the solar wind velocity and magnetic field, the tilt of the HCS is not the smallest during cycle 23, but larger than that during the previous two minima of cycles 21 and 22 for which the HCS tilt data are available (WSO website, http://wso.stanford.edu). Further, in contrast to the minima of cycles 21 and 22, the tilt angle during the deep minimum of cycle 23 changes very rapidly by about 25° in a span of about 20 solar rotations (see Figure 1(e)), during a period when sunspots were almost absent. This difference in solar/heliospheric parameters (sunspot number, solar wind velocity, interplanetary magnetic field strength, and the tilt of the HCS) during this peculiar minimum is expected to throw some light as regards the models for cosmic ray modulation. Thus, we have plotted the neutron monitor data as observed at Oulu station, during the decreasing and minimum solar activity phases of cycles 20, 21, 22, and 23 (Figure 1(f)). These periods correspond to the recovery of cosmic ray modulation cycles. In addition to differences in cosmic ray recovery during odd cycles (21, 23) as compared to even cycles (20, 22), we can see that during the deep minimum of cycle 23, the cosmic ray intensity reached the highest level, never recorded earlier in neutron monitor records (see Figure 1(f)).

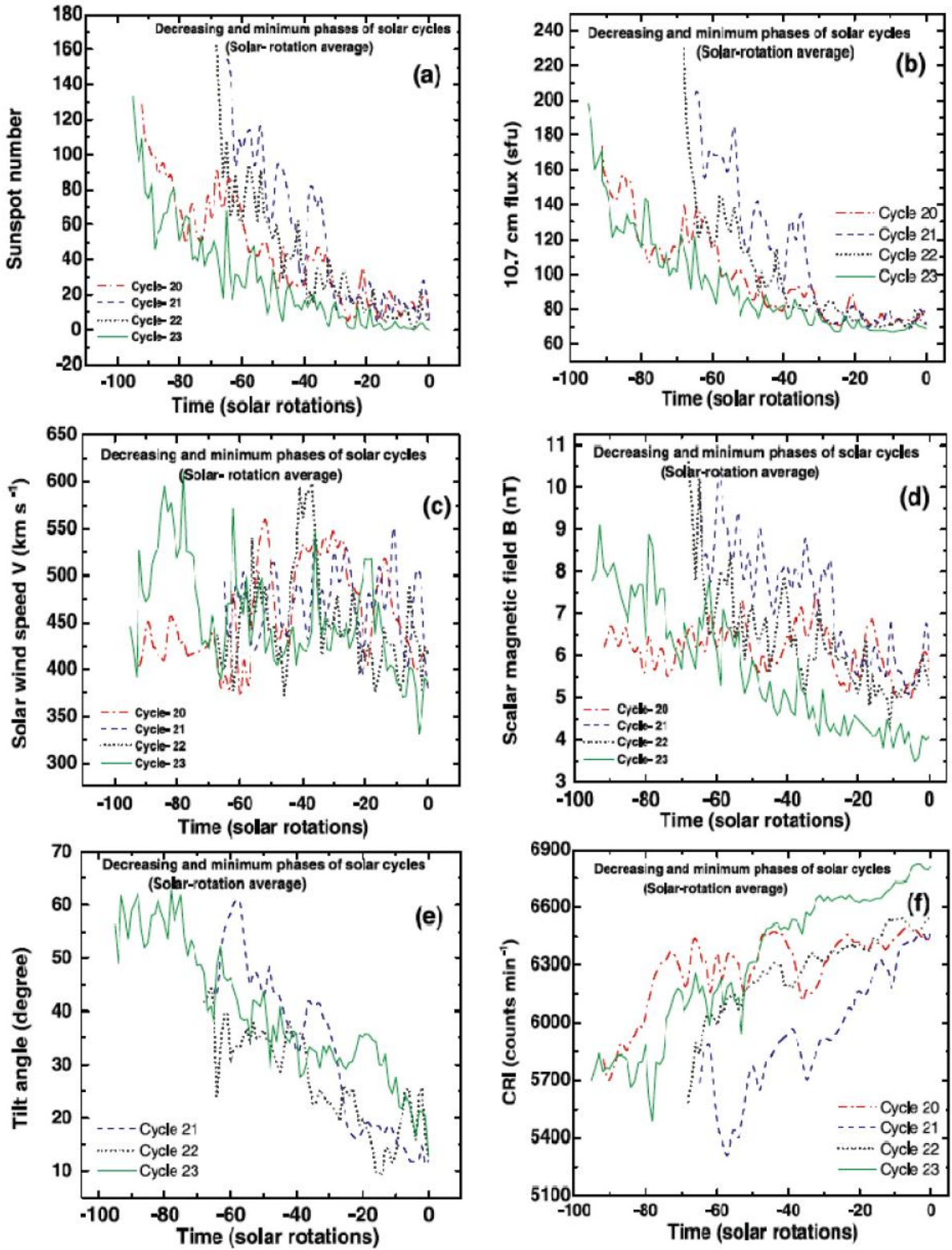

**Figure 1** Comparison of various parameters (27-day average) during the declining and minimum phases of solar cycles 20, 21, 22, and 23: (a) sunspot number, (b) 10.7 cm radio flux, (c) solar wind speed, (d) interplanetary magnetic field, (e) tilt angle of the heliospheric current sheet, and (f) cosmic ray intensity. On the time axis '0' corresponds to the time of the last rotation in each solar cycle, after which the solar activity (as seen in sunspot numbers) starts rising, marking the beginning of the next cycle.

## 2.2. Time Evolution and Best-Fit Curves during the Declining and Minimum Phases of Different Solar Cycles

The time evolution of two solar parameters (SSN, 10.7cm solar flux), three solar wind parameters ($V$, $B$, $\sigma_B$), the tilt angle ($\alpha$) of the HCS, and also the GCR intensity in the declining and minimum phases of cycles 20, 21, 22, and 23 were fitted with a polynomial ($y = A + B_1 x + B_2 x^2$). How good or bad is the fit is given by the values of the determination coefficient ($R^2$, the square of the correlation coefficient) in Table 1. In order to compare the decay of different solar and interplanetary parameters and look for differences during different cycles, these fitted polynomials for four cycles are plotted separately for two solar activity parameters (SSN, 10.7cm solar flux) [Figure 2a, 2b], two solar wind related parameters ($V$, $B$) [Figures (2c, 2d)], the tilt of the HCS ($\alpha$) [Figure 2e], and finally the GCR intensity [Figure 2f].

There was some controversy whether the odd numbered solar cycle 23 was evolving during its early phase, similar to the even numbered cycle 20 or the odd numbered cycle 21 (see, *e.g.,* Cliver and Ling, 2001; Dmitriev, Suvorova, and Veselovsky, 2002; Van Allen, 2003; Özgüç and Ataç, 2003; Singh, Singh, and Badruddin, 2008). Van Allen (2003) and Cliver and Ling (2001) reported that the early phase of cycle 23 resembles more those of cycles 19 and 21. On the other hand Özgüç and Ataç (2003) and Dmitriev, Suvorova, and Veselovsky (2002) reported that solar activity cycle 23 was evolving in a manner generally similar to cycle 20. Now, since the SSN data of the decay phase of all four solar cycles is available, it will be interesting to see the similarities/dissimilarities in the decay phases of these four cycles 20, 21, 22, and 23.

We observe that the sunspot numbers appear to decay more quickly during cycles 21 and 22 as compared to 20 and 23, to reach their lowest values. Moreover, there appears more similarlity in the time evolution of sunspot numbers between cycle 23 and cycle 20. The decay of sunspot numbers in cycle 21 appears more similar to cycle 22 (Figure 2a).

The time evolution of 10.7cm solar radio flux during the corresponding phases of solar cycles is similar to the sunspot numbers as expected. Similarity in the decay of solar radio flux in cycles 20 and 23, and that in cycles 21 and 22 is apparent as in the case of sunspot numbers. It is faster in the case of former two cycles as compared to the latter two (Figure 2b).

The solar wind speed during the declining and minimum phases of solar cycles 20, 21, and 22 do not decrease progressively as observed in the case of SSN and 10.7cm solar flux. However, although fluctuations are large and the polynomial fitting is poor (small correlation coefficients), the average behaviour of the time variation of the solar wind velocity can be approximated as, first increasing and then decreasing during this phase of cycles 20, 21, and 22 (Figure 2c). However, the time evolution of the solar wind speed during the same phase of cycle 23 is unique when compared to other three cycles, which are similar in the sense that the speed initially increases and then decreases. In contrast to the previous three cycles, in cycle 23 the solar wind speed can be fitted by an almost linearly decreasing curve. Thus the time variation of the solar wind speed is peculiar in cycle 23, during this phase as compared to the previous three solar cycles during the same phase of solar activity cycles.

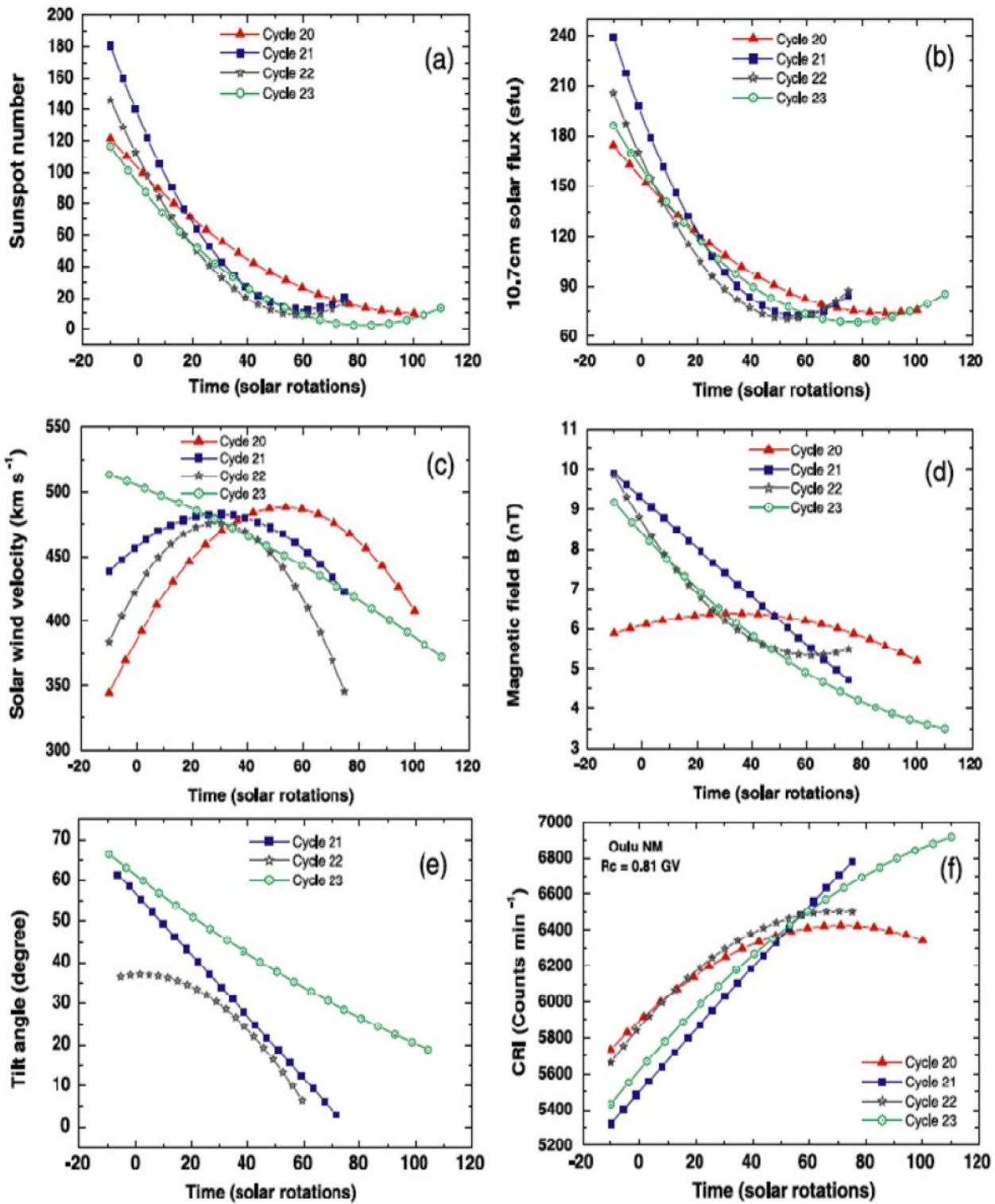

**Figure 2** Best-fit curves to the data shown in Figure 1.

**Table 1** Best fit coefficients $B_1$ and $B_2$ of the polynomial fit ($y = A + B_1x + B_2x^2$) to the time evolution of different parameters during the declining and minimum phases of solar cycles 20, 21, 22, and 23. $R^2$ is the determination coefficients.

| Solar cycle | SSN | | | 10.7 cm flux | | | V | | | B | | |
|---|---|---|---|---|---|---|---|---|---|---|---|---|
| | $B_1$ | $B_2$ | $R^2$ | $B_1$ | $B_2$ | $R^2$ | $B_1$ | $B_2$ | $R^2$ | $B_1$ | $B_2$ | $R^2$ |
| 20 | -1.8 ± 0.2 | 0.009 ± 0.002 | 0.82 | -1.84 ± 0.17 | 0.01 ± 0.001 | 0.84 | 3.83 ± 0.74 | -0.04 ± 0.008 | 0.25 | 0.02 ± 0.009 | $-2.7*10^{-4}$ ± $9.5*10^{-5}$ | 0.16 |
| 21 | -4.1 ± 0.44 | 0.03 ± 0.006 | 0.84 | -4.26 ± 0.38 | 0.04 ± 0.01 | 0.87 | 1.69 ± 1.12 | -0.03 ± 0.02 | 0.05 | -0.06 ± 0.02 | $-2.65*10^{-5}$ ± $3.28*10^{-4}$ | 0.65 |
| 22 | -3.4 ± 0.3 | 0.03 ± 0.004 | 0.86 | -3.61 ± 0.28 | 0.03 ± 0.00 | 0.86 | 3.51 ± 1.16 | -0.06 ± 0.02 | 0.22 | -0.11 ± 0.02 | $-8.77*10^{-4}$ ± $2.37*10^{-4}$ | 0.69 |
| 23 | -2.2 ± 0.15 | 0.01 ± 0.002 | 0.88 | -2.39 ± 0.13 | 0.02 ± 0.001 | 0.90 | -0.85 ± 0.68 | -0.003 ± 0.007 | 0.33 | -0.07 ± 0.008 | $-2.77*10^{-4}$ ± $7.83*10^{-5}$ | 0.87 |

| Solar cycle | $\sigma_B$ | | | $\Lambda$ | | | CRI | | |
|---|---|---|---|---|---|---|---|---|---|
| | $B_1$ | $B_2$ | $R^2$ | $B_1$ | $B_2$ | $R^2$ | $B_1$ | $B_2$ | $R^2$ |
| 20 | 0.01 ± 0.01 | $-2.5*10^{-4}$ ± $1.02*10^{-4}$ | 0.21 | -- | -- | -- | 14.8 ± 1.9 | -0.10 ± 0.02 | 0.62 |
| 21 | -0.02 ± 0.02 | $-4.57*10^{-4}$ ± $3.23*10^{-4}$ | 0.63 | -0.74 ± 0.14 | $-1.7*10^{-4}$ ± 0.00 | 0.88 | 17.8 ± 3.09 | -0.01 ± 0.05 | 0.90 |
| 22 | -0.09 ± 0.02 | $-4.75*10^{-4}$ ± $2.67*10^{-4}$ | 0.67 | 0.04 ± 0.13 | -0.005 ± 0.002 | 0.79 | 18.4 ± 1.54 | -0.13 ± 0.02 | 0.90 |
| 23 | -0.07 ± 0.01 | $-238*10^{-4}$ ± $7.7*10^{-5}$ | 0.86 | -0.52 ± 0.07 | $-0.001*10^{-4}$ ± $-6.7*10^{-4}$ | 0.87 | 18.4 ± 1.46 | -0.06 ± 0.01 | 0.93 |

The time variations of interplanetary magnetic field *B* during the same phase of solar cycles 20, 21, 22, and 23 are represented by the best-fit polynomials and are shown in Figure 2d. We observe that, in cycle 20, the interplanetary magnetic field *B* does not change much during this phase of solar cycle. It was expected to decrease during the decreasing sunspot activity. It is possible that, during the early years of observations (*e.g.,* early 1970's), the quality of magnetic field data was not good enough, as we see large data gaps in hourly data sets. From a comparison of the time behaviour of other three cycles (21, 22 and 23), we conclude that in even cycle 22, *B* appears to decrease more quickly after solar maximum, than in odd cycles 21 and 23. However, it becomes constant earlier in cycle 20 than in cycles 21 and 23. Moreover, the initial decay rate of *B* appears to be nearly the same in both odd cycles.

**Table 2** Decrease in the cosmic ray intensity against the changes in various parameters ($\Delta I/\Delta P$) and their correlation coefficients ($R$) during the declining and minimum phases of solar cycles 20, 21, 22, and 23.

| Solar cycle | | Phase | Sunspot number | | 10.7cm radio flux | | Magnetic field | | SD of magnetic field | | Solar wind velocity | | Tilt angle | |
|---|---|---|---|---|---|---|---|---|---|---|---|---|---|---|
| | | | $\Delta I/\Delta P$ | $R$ | $\Delta I/\Delta P$ | $R$ | $\Delta I/\Delta P$ | $R$ | $\Delta I/\Delta P$ | $R$ | $\Delta I/\Delta P$ | $R$ | $\Delta I/\Delta P$ | $R$ |
| 20 | $A>0$ | Declining and minimum | $-5.06 \pm 0.41$ | $-0.79$ | $-5.75 \pm 0.45$ | $-0.80$ | $-132.2 \pm 29.9$ | $-0.42$ | $-131.0 \pm 26.4$ | $-0.46$ | $-0.01 \pm 0.39$ | $-0.009$ | -- | ---- |
| 21 | $A<0$ | Declining and minimum | $-5.50 \pm 0.55$ | $-0.78$ | $-5.82 \pm 0.58$ | $-0.80$ | $-185.7 \pm 14.8$ | $-0.84$ | $-190.7 \pm 16.0$ | $-0.83$ | $-0.27 \pm 0.9$ | $-0.04$ | $-19.84 \pm 0.90$ | $-0.94$ |
| 22 | $A>0$ | Declining and minimum | $-5.42 \pm 0.27$ | $-0.93$ | $-5.75 \pm 0.27$ | $-0.93$ | $-146.8 \pm 9.60$ | $-0.88$ | $-125.8 \pm 10.4$ | $-0.83$ | $-0.71 \pm 0.46$ | $-0.19$ | $-14.9 \pm 0.78$ | $-0.92$ |
| 23 | $A<0$ | Declining and minimum | $-10.6 \pm 0.54$ | $-0.90$ | $-10.37 \pm 0.56$ | $-0.89$ | $-241.6 \pm 7.6$ | $-0.96$ | $-256.4 \pm 9.7$ | $-0.94$ | $-4.39 \pm 0.49$ | $-0.67$ | $-27.8 \pm 1.09$ | $-0.93$ |

The HCS separates the two oppositely directed magnetic polarity hemispheres of the heliosphere (Jokipii and Thomas, 1981). The tilt angle of the HCS shows a variation with solar activity and exhibits a periodicity of about 11years. An inclined current sheet has a significant effect on the global heliospheric field and on the drift motions of cosmic ray particles. Thus, the tilt of the HCS has become a prime indicator of solar activity from thet point of view of particle drifts, and the wavy HCS has turned out to be one of the most significant physical effect in the modeling of cosmic ray modulation.

The intensity of cosmic ray particles entering the heliosphere is modulated as they travel through the heliospheric magnetic field embedded in the solar wind (Parker, 1965; Rao 1972; Venkatesan and Badruddin, 1990; Potgieter *et al.*, 2001). The large-scale IMF consists of the so-called Parker spiral, and the opposite magnetic hemispheres are divided by a thin HCS. The polarity of the solar polar magnetic fields and the heliosphere changes around the period of solar activity maximum. In the seventies (1971-1979) and nineties (1991-1999), for example, the field is directed outward in the northern hemisphere and inward in the southern hemisphere. In this configuration, referred to as *A*>0, positively changed GCR particles drift inward through the poles and then downward from the poles towards the HCS (near the equator). In the opposite polarity configuration (when the field is directed inward in the northern hemisphere and outward in the southern heliosphere) as, for example, in the sixties (1961-1969), eighties (1981-1989) and (2001-?), referred to as *A*<0, GCR particles drift inward along the HCS and then upward toward the poles (Kota and Jokipii, 1983; Gupta and Badruddin, 2009). Thus one expects that incoming GCR particles will be affected differently by the drift effects between the two magnetic configurations *A*>0 and *A*<0.

The evolution of the tilt angle during the declining and minimum phases of odd (21 and 23) solar cycles can be best fitted by an almost linear curve. As shown in Figure 2e, the rate of decay of the tilt angle during this period is slower in cycle 23 as compared to cycle 21 (both are $A<0$ epochs) (see also McDonald, Webber, and Reames, 2010). However, during the same phase in cycle 22 ($A>0$), the decay is not similar as in cycles 21 and 23. Cliver, Richardson, and Ling (2011) reported that the evolution of the tilt angle appears to be systematically different in even- and odd-numbered cycles.

The increase in GCR intensity during the same phases of solar activity cycles 20, 21, 22, and 23 and the fitted polynomial ($y = A + B_1 x + B_2 x^2$) are shown in Figure 2f for comparison. The difference between even (20, 22) and odd (21, 23) cycles is apparent here; 'broad' maxima in GCR intensity during even cycles (20, 22) as compared to 'peaked' maxima during odd cycles (21, 23). This difference in pattern, expected from the prediction of drift theory of cosmic ray modulation, is well represented here by the fit using a polynomial of second order. These fits appear good, as is apparent from the values of the determination coefficients (see Table 1).

## 2.3. Comparison between GCR Intensity Variation and Solar/Interplanetary Parameters during the Declining and Minimum Phases: Best-Fit Approach to Individual Cycles

None of the parameters (SSN, 10.7 cm flux, $V$, $B$, and ) follow the pattern (of course in anti-phase) of time evolution of cosmic ray intensity (CRI) during the declining and minimum phases of all four cycles 20, 21, 22, and 23. However, individual solar/interplanetary parameters may track well (in anti-phase) the time evolution of GCR intensity during some of the cycles. This can be inferred from Figures 1(a-f) and 2(a-f) and from the correlation coefficients given in Table 2.

During cycle 20, out of five solar and interplanetary parameters (SSN, 10.7cm solar flux, $V$, $B$, and $_B$) only solar activity parameters (SSN and 10.7cm solar radio flux) appear well (anti) correlated with CRI during the declining phase (and not during the minimum) of this solar cycle (see Tables 3 and 4). During odd cycle 21, the increase in CRI during the declining phase correlates well with the decrease in the tilt angle ( ) ($R = -0.84$) and also correlates with the field amplitude ($B$) better than in cycle 20 ($R = -0.58$). During even cycle 22, solar parameters (SSN, 10.7cm solar flux), and the interplanetary parameters ($B$, $_B$) correlate well, in anti-phase, with CRI, as evident from the values of $R$ in Table 3.

In Figures 3 (a, b) we have plotted the best fit curves, exclusively for cycle 23, to compare the time evolution of various parameters (SSN, 10.7cm flux, , $V$, $B$ and $_B$) with CRI in the declining and minimum phases of cycle 23, as this period is of special interest. We can see that in the declining and minimum phases of cycle 23, in addition to $B$, $_B$, and , the solar wind velocity ($V$) too appears to track and correlate well with the change in CRI (see also Tables 2, 3, and 4). Note that, in Figures 3(a, b), the scale for CRI is inverted for comparison.

**Table 3** Decrease in the cosmic ray intensity against the changes in various parameters ($\Delta I/\Delta P$) and their correlation coefficients ($R$) during the declining phases of solar cycles 20, 21, 22, and 23.

| Solar cycle | | Phase | Sunspot number | | 10.7 cm radio flux | | Magnetic field | | SD of magnetic field | | Solar wind velocity | | Tilt angle | |
|---|---|---|---|---|---|---|---|---|---|---|---|---|---|---|
| | | | $\Delta I/\Delta P$ | $R$ | $\Delta I/\Delta P$ | $R$ | $\Delta I/\Delta P$ | $R$ | $\Delta I/\Delta P$ | $R$ | $\Delta I/\Delta P$ | $R$ | $\Delta I/\Delta P$ | $R$ |
| 20 | $A>0$ | Declining | -5.50 ± 0.65 | -0.73 | -5.80 ± 0.87 | -0.73 | -93.4 ± 45.9 | -0.25 | -122.2 ± 44.2 | -0.33 | 0.06 ± 0.44 | 0.02 | --- | --- |
| 21 | $A<0$ | Declining | -2.54 ± 0.80 | -0.46 | -2.81 ± 0.77 | -0.51 | -119.6 ± 27.9 | -0.58 | -101.0 ± 34.0 | -0.44 | 1.19 ± 0.80 | 0.24 | -21.03 ± 2.21 | -0.84 |
| 22 | $A>0$ | Declining | -4.69 ± 0.38 | -0.89 | -4.81 ± 0.30 | -0.93 | -123 ± 13.6 | -0.82 | -111.2 ± 18.6 | -0.69 | -0.70 ± 0.47 | 0.23 | -15.58 ± 1.53 | -0.85 |
| 23 | $A<0$ | Declining | -9.1 ± 0.69 | -0.85 | -8.71 ± 0.69 | -0.84 | -230.2 ± 11.5 | -0.93 | -238.6 ± 14.7 | -0.89 | -3.92 ± 0.54 | -0.67 | -27.3 ± 1.56 | -0.91 |

**Table 4** Decrease in the cosmic ray intensity against the changes in various parameters ($\Delta I/\Delta P$) and their correlation coefficients ($R$) during the minimum phases of solar cycles 20, 21, 22, and 23.

| Solar cycle | | Phase | Sunspot number | | 10.7 cm radio flux | | Magnetic field | | SD of magnetic field | | Solar wind velocity | | Tilt angle | |
|---|---|---|---|---|---|---|---|---|---|---|---|---|---|---|
| | | | $\Delta I/\Delta P$ | $R$ | $\Delta I/\Delta P$ | $R$ | $\Delta I/\Delta P$ | $R$ | $\Delta I/\Delta P$ | $R$ | $\Delta I/\Delta P$ | $R$ | $\Delta I/\Delta P$ | $R$ |
| 20 | $A>0$ | Minimum | -0.51 ± 0.82 | -0.12 | -1.93 ± 1.39 | -0.27 | -33.6 ± 9.5 | -0.57 | -24.7 ± 7.6 | -0.54 | -0.30 ± 0.15 | -0.37 | ---- | ----- |
| 21 | $A<0$ | Minimum | -3.85 ± 4.27 | -0.18 | -4.49 ± 8.37 | -0.11 | -120.8 ± 50.4 | -0.43 | -139.2 ± 32.1 | -0.65 | -1.37 ± 0.60 | -0.41 | -26.37 ± 4.66 | -0.75 |
| 22 | $A>0$ | Minimum | -4.89 ± 1.47 | -0.55 | -9.09 ± 2.5 | -0.59 | -69.9 ± 22.9 | -0.51 | -45.1 ± 16.1 | -0.48 | -0.49 ± 0.43 | -0.21 | -8.65 ± 1.73 | -0.71 |
| 23 | $A<0$ | Minimum | -8.45 ± 3.78 | -0.41 | -8.78 ± 5.27 | -0.32 | -164.7 ± 36.3 | -0.66 | -164.3 ± 41.3 | -0.61 | -1.11 ± 0.16 | -0.80 | -10.2 ± 0.83 | -0.92 |

## 2.4. Polarity-Dependent Effect in the Variability of GCR Intensity: Quantitative Relation among Different Parameters

Figures 1 and 2 give a qualitative comparison of the variability of various solar and interplanetary parameters, and cosmic ray intensity during the same phases of solar cycles 20, 21, 22, and 23. However, in order to understand the mechanism for cosmic ray modulation, one also needs to study the relationship between the variability of cosmic ray intensity with solar and interplanetary parameters during different polarity states of the heliosphere ($A>0$ and $A<0$). For this purpose we have considered the values of various parameters averaged over one solar rotation and studied the cross correlation plots between CRI and different solar and interplanetary parameters during the declining and minimum phases of solar cycle 20 ($A>0$), 21 ($A<0$), 22 ($A>0$), and 23 ($A<0$).

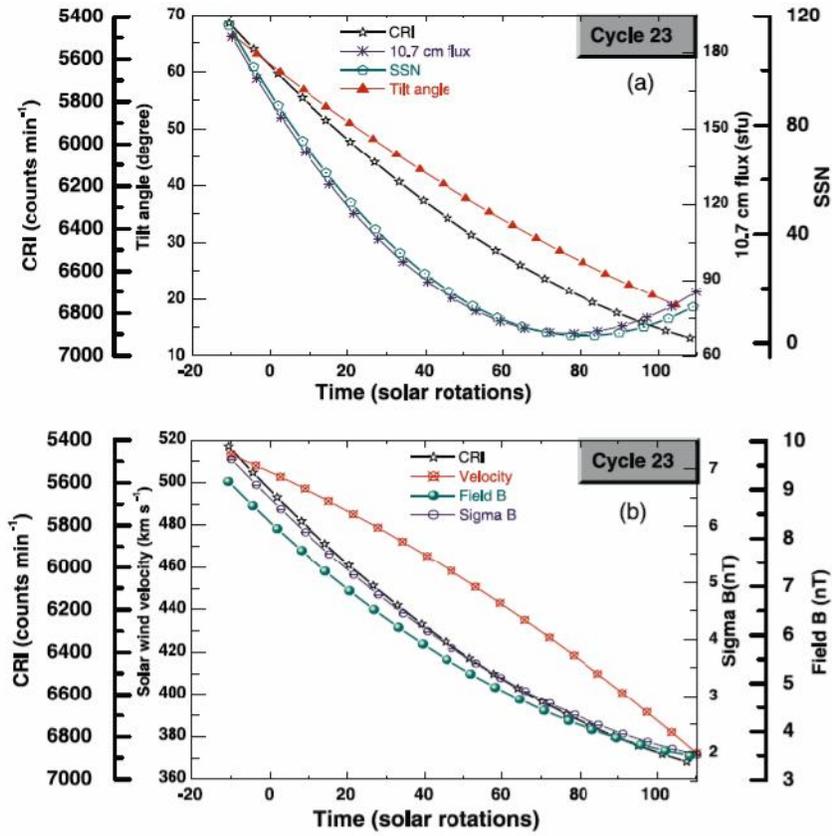

**Figure 3** Best-fit curves for (a) the sunspot number (SSN), 10.7cm flux, and HCS tilt angle in comparison with CRI, and (b) solar wind velocity, interplanetary field *B*, and $\sigma_B$ in comparison with CRI, during the declining and minimum phases of solar cycle 23. Note that the scale is inverted for CRI.

We have first determined the time lag between CRI and various solar/ interplanetary parameters during the declining and minimum phases of solar cycles 20, 21, 22, and 23. For determining the time lag between CRI and solar/interplanetary parameters (SSN, 10.7cm flux, *V*, *B*, and $\sigma$), we have calculated the correlation coefficients between the two by introducing successively the time lags of 0 to 20 solar rotations. From the optimum values of correlation coefficients between CRI and individual parameters, the most probable time lag for different parameters are inferred. Then the quantitative estimates of the rate of change in GCR intensity with different solar and interplanetary parameters ($\Delta I/\Delta P$), wherever applicable, have been done by linear regression analysis after introducing respective time lags. The slopes of the linear fit between rotation-averaged CRI and corresponding solar/interplanetary parameters ($\Delta I/\Delta P$) so obtained, and respective correlation coefficients are listed in Table 2. The GCR intensity increases at almost an equal rate with the decay in SSN and 10.7cm solar flux during even cycles 20 and 22. However, this rate is much faster during cycle 23, almost double the rate during the previous three cycles 20, 21, and 22. The correlation between CRI and solar/interplanetary parameters is particularly good in cycle 23.

Excluding cycle 20, where the correlation between CRI and interplanetary field (*B*) and standard deviation of vector field ($\sigma_B$) and is poor ($R<0.5$), we find that the rate of increase in CRI with the decrease in field *B* and $\sigma_B$ is significantly higher during the *A*<0 epochs (cycles 21 and 23) as compared to the *A*>0 epochs (cycles 20 and 22). As regards the solar wind speed (*V*), the correlations between CRI and V are very poor during this phase of cycle 20 (*R*= -0.009), 21 (*R* = -0.04), and 22 (*R*= -0.19). However, the correlation is good (R= - 0.67) in cycle 23.

It was shown (*e.g.,* Kota, 1979; Jokipii and Thomas, 1981) that the current sheet would play a more prominent role in the *A*<0 state when positively charged GCRs enter the heliosphere along the helio-equator and would interact with the HCS. Because the particles enter over the poles in the *A*>0 state, they rarely encounter the current sheet on their inward journey, and the GCR particle density is thus relatively unaffected by the current sheet in this state. Once it was realized that gradient and curvature drifts should play an important role in modulation, Jokipii and Thomas (1981) and Kota and Jokipii (1983) identified the inclination of the heliospheric current sheet as a key parameter for the models of GCR modulation. Correlation between the tilt angle and CRI has been obtained by Smith and Thomas (1986), Webber and Lockwood (1988), Smith (1990), and Badruddin, Singh, and Singh (2007). In accordance with the drift theory, the slope of these correlations depends on the polarity of the cycle, with CRI being less sensitive to changes in the tilt angle during the positive cycles (Fluckiger, 1991). They showed that there was a general inverse correlation between the CRI at the earth and the tilt of the HCS, and that the inverse relation was more pronounced during *A*<0 than *A*>0.

In agreement with earlier findings, our results indicate that the response of the decrease in the tilt angle to the GCR intensity is faster during *A*<0 (cycles 21 and 23) than in *A*>0 (cycle 22) as can be seen from the values of $\Delta I/\Delta P$ (see Table 2). Another observation of special mention is that, during the declining and minimum phases, the correlation between CRI and interplanetary parameters (*B*, $\sigma_B$, and *V*) is best during cycle 23 as compared to the previous three cycles 20, 21, and 22.

To see whether the relationship, mentioned above, between CRI and various solar and interplanetary parameters, including the tilt angle, persist during the minimum and the declining phases separately or it is influenced by a single phase (declining plus minimum), we have done similar analysis, separately during the declining and minimum phases of all four solar cycles 20, 21, 22, and 23. In Table 3, the relationship between CRI and various solar and interplanetary parameters during the declining phase is summarized. The difference in the increase rate of GCR intensity between the *A*<0 and *A*>0 periods, as observed during the declining and minimum phases (seen in Table 2) is so not obvious here. However, the rate of GCR intensity decrease is correlated with interplanetary parameters (*B*, $\sigma_B$, *V*, and $\alpha$) highly in the declining phase of cycle 23 when compared with the the same phase in cycle 20, 21, and 22. The correlations are also comparatively better in cycle 23. The correlation between CRI and solar wind *V* can be treated as significant during the declining phase of this last cycle only.

However, when studying exclusively the minimum phases of cycles 20, 21, 22, and 23 (Table 4), the GCR intensity increases with interplanetary parameters (*B*, $\sigma_B$, *V*, and $\alpha$) faster during *A*<0 as compared to *A*>0 periods. Further the correlation is particularly high with V (*R* = 0.80) and $\alpha$ (*R*= -0.92) during the unusual minimum of cycle 23.

The correlation coefficients and slopes obtained from a linear fit during (a) the declining and minimum phases, (Table 2) (b) the declining phase alone (Table 3), and (c) the minimum phase alone (Table 4) of these four solar activity cycles are tabulated. Looking at the rate of change of CRI with different parameters ($\Delta I/\Delta P$) and correlation coefficients ($R$), we conclude that in general the CRI decreases at a faster rate with respect to the increase in $B$ and the HCS tilt in the $A<0$ epoch. This trend is more clearly evident during the minimum phase of solar cycles. A similar trend in the decrease rate of CRI is also seen with respect to the solar wind velocity, although the correlations in this case are poor. Further, there are stronger correlations between CRI and the tilt of HCS during the $A<0$ epochs. Another observation of special mention is that the correlation coefficients between CRI and the magnetic field strength ($B$), solar wind speed ($V$), and tilt angle ($\alpha$) of the HCS are the highest during cycle 23 ($A<0$) compared to all the other previous cycles, both in the declining as well as the minimum phases of solar cycles. Although the above mentioned results can broadly be extracted from our analysis; these results are not strictly steady for all cycles and insignificant in some cases. Wibberenz, Richardson, and Cane (2002) reported that for the low $B$ ($\leq$ 6nT) period in the $A>0$ epochs the GCR intensity decreases more slowly as B increases than in the case for the $A<0$ epochs.

## 2.5. Deep Solar Minimum of Cycle 23

The recent minimum of cycle 23 is the most completely observed minimum as there are a large number of ground based and space-based observatories now operating worldwide. Moreover, the recent peculiar solar minimum of cycle 23 has been unusually long and deep. In Figure 4 we have plotted the values of solar (SSN, 10.7cm solar flux) and interplanetary ($B$, $\sigma_B$, $V$, and $\alpha$) parameters averaged over one rotation, along with the GCR intensity (in percent) during the minimum phase of cycle 23 (shaded area). For comparison, we have also plotted these parameters a few solar rotations before and after the minimum. A record high GCR intensity and record low values of $B$, $\sigma_B$, and $V$ are the highlight of Figure 4 during the minimum of cycle 23. During most of the time period in the cycle 23 minimum, sunspots are almost absent and essentially there is little or no variability in solar activity as seen from sunspot numbers. As regards the interplanetary parameters $V$ and $B$, although they reached a record low level during this period, they are variable in time, particularly the solar plasma speed. For the point of view of cosmic-ray modulation, it is interesting to note that, although the intensity of solar polar field was very low and almost constant (Jian, Russell, and Luhmann, 2011), the tilt of HCS was changing rapidly; a decrease of about $25^o$ during this peculiar minimum itself. Moreover, the tilt angle is not at minimum as compared to the behaviour of the tilt angle during the minima of cycles 21 and 22. However, the cosmic ray intensity reached the record highest level in the whole era of neutron monitor observations. The increase in GCR intensity to a record maximum level is an indication that this minimum had an extraordinary effect on the properties of the magnetic structure shielding the earth to allow such an increase in GCR intensity (White *et al.,* 2011). To highlight some of the points specific to cycle 23, regarding the cosmic ray modulation, in Figure 5, we have plotted the scatter diagram, and best fit linear curve, exclusively during the minimum. We find that, during this peculiar cycle 23 minimum, CRI shows poor correlation with sunspot number ($R = -0.41$), better correlation with magnetic field ($R= -0.66$), still better correlation with solar wind speed ($R = -0.80$) and much better correlation with the tilt angle of HCS ($R = -0.92$) among all the parameters considered. In other words, the correlations of GCR intensity, during the cycle 23 minimum, are progressively higher with respect to $B$, $V$, and  (see Figure 5 and Table 4). These values during cycle 23 are highest when compared with the same parameters in the same phase in cycles 20, 21, and 22.

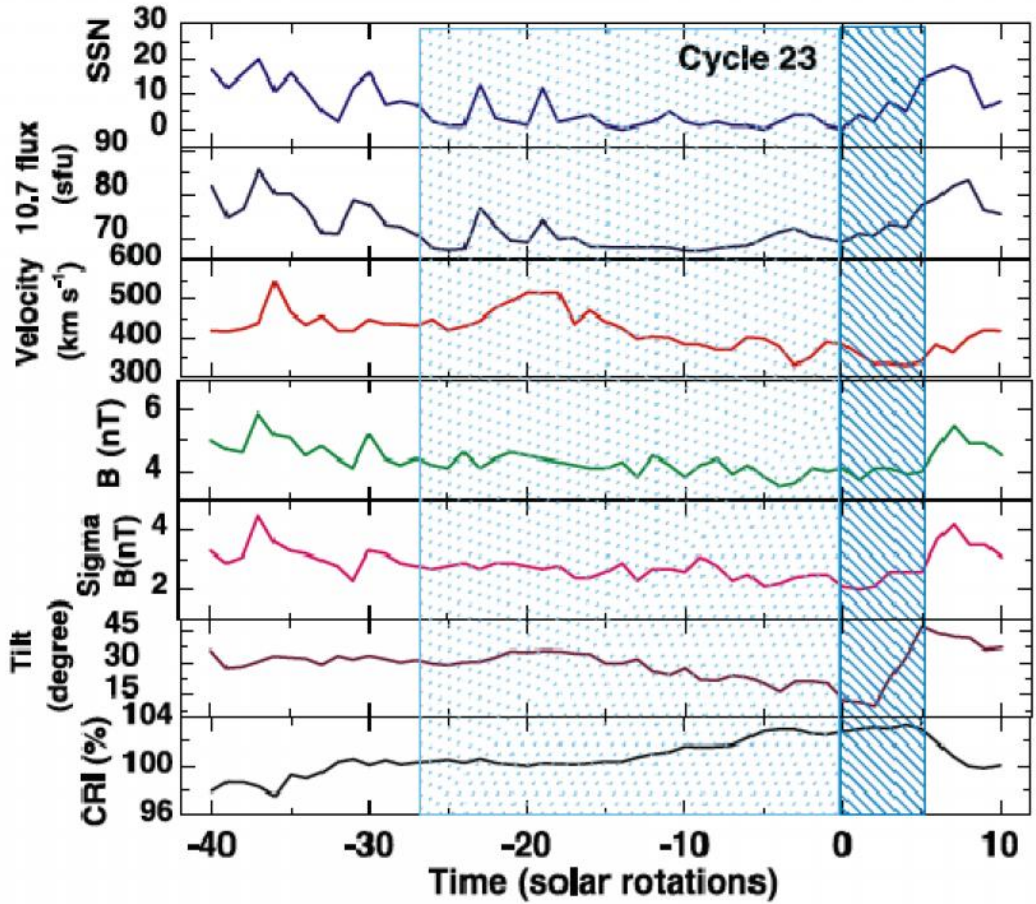

**Figure 4** The values averaged over one solar rotation of sunspot numbers, 10.7cm solar flux, solar wind velocity V, interplanetary magnetic field B, Sigma B, the tilt angle of the heliospheric current sheet, and Galactic cosmic ray intensity during the minimum phase of solar cycle 23 (shaded area by dots). Five solar rotation periods after the end of minimum '0' are marked (slanted lines) highlighting the period when the cosmic ray intensity is at the highest level.

The transport equation of cosmic-ray particles has terms governing GCR modulation in the heliosphere as diffusion, convection, curvature/gradient drift, and adiabatic deceleration in the heliosphere. The transport problem aims at an understanding of all those processes determining the propagation of GCRs in the heliosphere and has been investigated with great effort. Measurements of anomalous cosmic ray (ACR) and galactic cosmic ray (GCR) intensities throughout the heliosphere have been used to examine the role of drifts, convection, and diffusion of GCR particles in the heliosphere over the past several solar cycles. During the past three decades, research on the GCR modulation has been primarily focused on understanding and specifying the roles of two of these terms -- diffusion and curvature/gradient drift. Interplanetary field strength ($B$) and the tilt angle ( ) are widely used to characterize diffusion and drift effects on cosmic ray intensity, respectively. In the current paradigm for the modulation of GCRs, diffusion is considered to be the dominant process during solar maxima while drift dominates at solar minima.

Cosmic ray modulation during the recent unusual solar minimum of cycle 23 was studied using the neutron monitor data (Gushchina *et al.,* 2009; Moraal, Stoker, and Kruger, 2009; Moraal and Stoker, 2010; Ahluwalia and Ygbuhay, 2010; Cliver, Richardson, and Ling, 2011; Jian, Russell, and Luhmann, 2011) and by utilizing both anomalous and galactic cosmic ray data (Leske *et al.,* 2009, 2011; Mewaldt *et al.,* 2010; McDonald, Webber, and Reames, 2010). However, the physical processes suggested by different authors, responsible for the unusual GCR intensity observed during this particular minimum, are in some cases quite different.

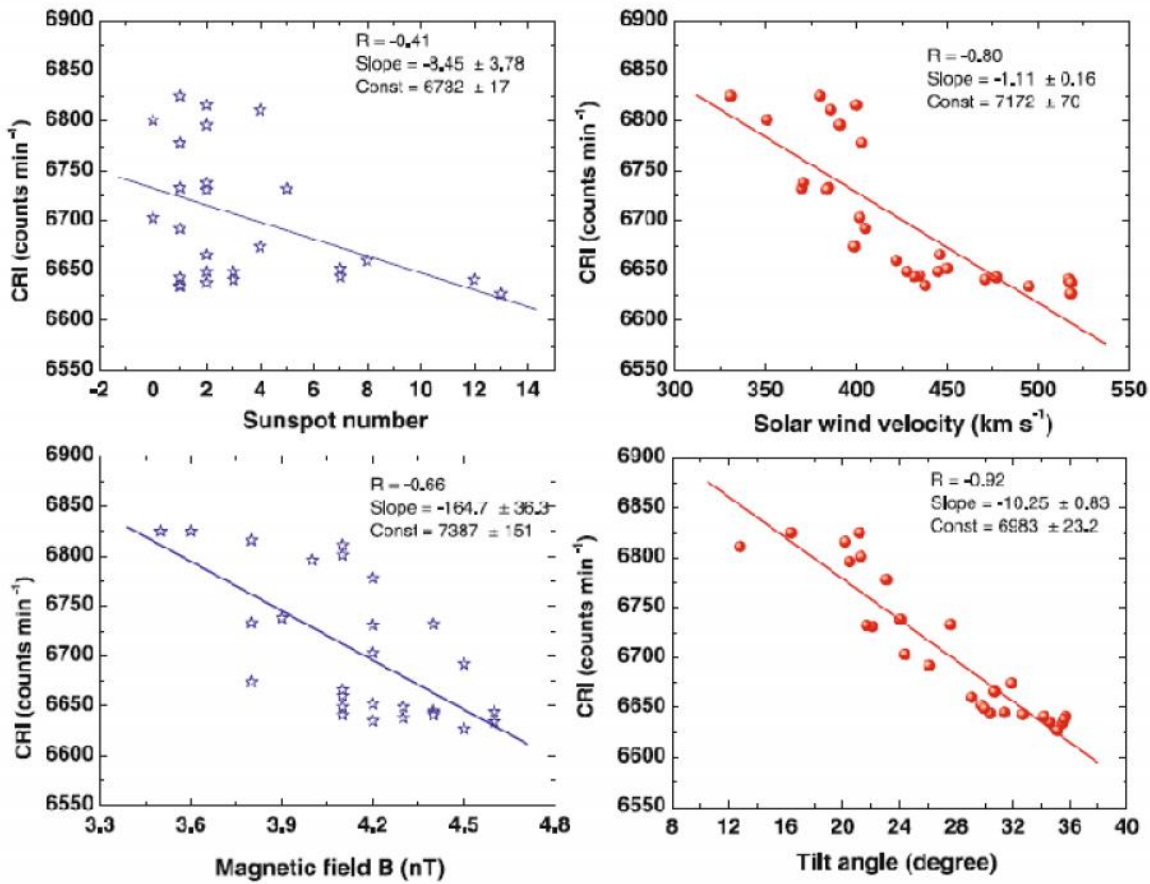

**Figure 5** Scatter plots between the cosmic ray intensity (CRI) and solar and interplanetary parameters, together with best-fit lines, exclusively during the minimum phase of solar cycle 23.

The solar minimum of cycle 23 has been marked by a prolonged and continuing period of very low solar activity. During 2008 there were no sunspots observed on 266 days of the year's 366 days (73%), during 2009 the spotless days were 274 of 365 days (75% spotless days). These represent the deepest minimum compared with the records of the 20th century (Hady, 2010). During this unusual minimum, the polar magnetic fields of the Sun are about half that of the other recent solar minima (Wang, Robbecht, and Sheeley, 2009), and the average tilt angle of HCS remained above $10°$, appreciably higher than observed over the previous two solar minima. The interplanetary magnetic field is 28% lower than during the previous four solar minima (Smith and Balogh, 2008). McComas *et al.* (2008), using Ulysses data, have reported a 20% decrease in the solar wind pressure compared to the cycle 22 minimum, which will have an effect on the size and structure of our heliosphere (see, *e.g.,* McDonald, Webber, and Reames, 2010).

It has been reported (Leske *et al.,* 2009) that the behaviour of the ACR and GCR intensity is unusual during the cycle 23 minimum, and that not only are the differences between $A>0$ and $A<0$ cycles but also there can be considerable variability between different $A<0$ cycles. Observations from Ulysses show that the average dynamic pressure in the solar wind has had a significant long term decline (McComas *et al.,* 2008) and that the heliospheric magnetic flux has decreased (Smith and Balogh, 2008); such global heliospheric changes would be expected to affect turbulence levels and drifts velocities, possibly altering the relative importance of diffusion versus drift effects between different solar cycles. Moraal, Stoker, and Kruger (2009) suggested that, although the solar activity parameters during the most recent solar minimum are quite different from the previous solar minima, the control of these parameters over the cosmic ray modulation is still according to the current paradigm. More recently, Moraal and Stoker (2010) reached at the conclusion that during the 2009 minimum, solar activity parameters were significantly different from the previous solar minima: The Sun was much quieter during this minimum, and the heliospheric magnetic field was 28% weaker than during the other recent minima. Both of these parameters (lower solar activity and weaker magnetic field) imply a higher cosmic ray diffusion coefficient, which provides a natural explanation for the higher galactic cosmic ray intensities that were observed.

Ahluwalia and Yghuhay (2010) found a good inverse correlation between the neutron monitor data in 2006-2009 and the monthly average of the interplanetary magnetic field strength $B$. It is well known that cosmic ray intensities are inversely correlated with $B$ (Burlaga and Ness, 1998; Cane *et al.,* 1999). In the efforts of theoretical modeling of cosmic rays in the heliosphere, the parallel diffusion coefficient ($K$) is assumed to be proportional to $1/B$ (Jokipii and Davila, 1981; Reinecke, Moraal, and McDonald, 2000). Further, the diffusion coefficient perpendicular to the magnetic field ($K$) is often assumed to scale as the parallel diffusion coefficient (Ferreira and Potgieter, 2004). In addition, Jokipii and Kota (1989) suggested that the drift velocities of GCR increase with decreasing $B$ (see Mewaldt *et al.,* 2010).

In contrast to the previous minima, solar minimum 23 has the smallest SSN, smaller 10.7cm radio flux, slower solar wind, much weaker solar and interplanetary field, and more warped HCS. Thus, there may be several factors (see McDonald, Webber, and Reames, 2010; Mewaldt *et al.,* 2010) that could cause the record high cosmic ray maximum during this minimum; weaker IMF, reduced IMF turbulence, slower solar wind, and weaker solar wind dynamic pressure than in the previous minima. In addition, the HCS, although not flatter than in the previous minima, is much flatter than during the other phases of cycle 23. As the warp of HCS declined in 2009, cosmic rays finally obtained a more direct inward access than along the earlier wavy HCS (Jian, Russell, and Luhmann, 2011).

McDonald, Webber, and Reames (2010) outlined two major effects that produce the unusual time histories during the cycle 23 minimum compared to those of cycles 19 and 21; the weaker interplanetary magnetic field and the slower approach and higher value of the tilt angle. They suggest that reductions in the solar wind speed, density, and the interplanetary field *B* could lead to larger values of the diffusion coefficients and corresponding changes in the structure of the heliosphere.

Mewaldt *et al.* (2010) observed that during the extended solar minimum in 2008-2010, there was a 1.5 year period (October 2008 through March 2010) during which the intensity of heavy (He to Fe) GCRs in the energy range from 70 to 450 MeV/nucleon was higher than the intensities during the previous four solar minima, and most neutron monitors were also at record high levels (*e.g.,* Ahluwalia and Yghuhay, 2010). The 2008-2010 solar minimum was unusual in several ways, including greatly reduced IMF strength and a prolonged decrease in the interplanetary turbulence level. According to Mewaldt *et al.* (2010) this combination produced a sustained increase in the estimated cosmic ray parallel mean free path and also increased GCR drift velocities.

Leske *et al*. (2011) argued that some of the factors likely to be important for heliospheric energetic particle intensities include a long duration for the minima, a weak interplanetary magnetic field, reduced turbulence, a drop in the solar wind dynamic pressure, and a very slowly declining HCS tilt angle, probably associated with the Sun's weaker polar magnetic fields. They suggested that although the reduced solar wind turbulence and magnetic flux would result in less modulation, this would have been partially compensated by the higher HCS tilt angle, which would increase the level of modulation. They further suggested that, for a given tilt angle, both GCR and ACR intensities are enhanced in the current $A<0$ cycle compared to the last $A<0$ cycle, but the enhancement in the GCR intensity is much more than in the ACR intensity.

However, Cliver, Richardson, and Ling (2011) suggested that observations during the recent solar minimum challenge the pre-eminence of drift at such time and that the tilt angle variation does not drive modulation at this solar minimum and that the record high cosmic ray intensities observed in 2009 resulted from a reduction in *B*, rather than from a reduction in the tilt angle. These authors supported the idea that the tilt angle variation does not drive modulation at a solar minimum as is commonly thought to be the case (*e.g.,* Potgieter and Le Roux, 1994; Jokipii and Wibberenz, 1998; Ferreira and Potgieter, 2004).

It is known (Jian, Russell, and Luhmann, 2011) that the high-speed streams (HSS) were frequently observed during the cycle 23 minimum. Cliver, Richardson, and Ling (2011) suggested that GCR intensity is weakly anti-correlated with the contribution from the HSS to magnetic field *B*.

It has been suggested earlier that the response to the GCR intensity modulation from the changing tilt angle during different solar magnetic cycles is not the same (*e.g.,* Webber and Lockwood, 1988; Smith, 1990; Badruddin, Singh and Singh, 2007). Further, the response of the solar wind speed against the 27-day variation of cosmic rays during the minima of different solar cycles may be different (*e.g.,* Singh and Badruddin, 2007; Modzelewska and Alania, 2011). Although the tilt of the HCS is rapidly decreasing during the minimum of cycle 23, the tilt angle is not at its minimum as compared to the tilt angle during the minima of cycle 21 and 22. It is known that the diffusion coefficient of GCR particles increases with decreasing interplanetary field magnitude $B$, and in addition, the drift velocities of GCR also increase with decreasing $B$ (see Mewaldt *et al.* (2010) and references therein).

We have shown in this work that the GCR intensity is better (anti-) correlated ($R = -0.80$) with the velocity of the solar wind than with the interplanetary magnetic field ($R = -0.66$) during the minimum period of cycle 23. Moreover, the correlation between GCR intensity and the tilt angle of the HCS is good ($R = -0.92$) during this particular minimum.

In view of the results discussed above, we suggest that it is not diffusion or drift alone, but convection in the solar wind is the most likely additional effect responsible for the record high GCR intensity observed during the particular and deep minimum of solar cycle 23.

## 3. Conclusions

Our study of the declining and minimum phases of solar cycles 20 ($A>0$), 21 ($A<0$), 22 ($A>0$), and 23 ($A<0$), with particular attention during the deep minimum of solar cycle 23, leads to the following conclusions.

- We observe, in agreement with previous workers, a slowly varying SSN and $B$, reaching to a record low level at the end of solar cycle 23. As regards the solar wind velocity, although it reached a record low level, its decrease towards the minimum is not exactly similar to SSN and $B$. The tilt of the HCS was decreasing slowly till about 20 rotations before the end of the minimum, after which it decreased faster to a value, which is still higher, compared to the smallest tilt angle during the previous minima. However, GCR intensity has reached a record high level in space era.

- The decay of solar activity (SSN and 10.7cm solar flux) to the minimum activity level in cycle 21 is similar to cycle 22, while of cycle 23 the decay shows more similarity with cycle 20.

- During the decay phase of solar cycles the behavior of the solar wind velocity is unique during cycle 23, decreasing with time as compared to cycles 20, 21, and 22. During this phase of the solar cycle, the speed of the solar wind bears some similarity with the other three cycles 20, 21, and 22, first increasing then decreasing, although the polynomial fit is not so good as evident from the values of the determination coefficients ($R^2$). This behaviour is probably due to the presence of high speed streams from coronal holes during the decreasing phase of the solar cycle.

- The decay of the magnetic field $B$ bears resemblance to solar activity (SSN, 10.7cm solar flux) during cycles 22 and 23 at least, during the declining phase.

- The evolution of the tilt angle during the same phase of solar cycles are different in even (22) and odd (21, 23) cycles. However, decay in the tilt angle is slower during cycles 23 as compared to cycle 21.

- The evolution of GCR intensity is different in odd (21, 23) and even (20, 22) cycles as expected from the drift theory.

- GCR intensity appears to decrease at a faster rate with the increases in all three interplanetary parameters (*B*, *V*, and tilt angle) during the *A*<0 epoch as compared to the *A*>0 epoch. This difference in rates ( *I*/ *P*) is particularly high during the solar minima of cycles 20, 21, 22, and 23. As compared to the other solar cycles, the correlations between the GCR intensity and parameters *B*, *V*, and tilt angle are the highest during the deep minimum of cycle 23.

- During the last five rotations, when GCR intensity was at its record highest level, (a) SSN was increasing after an extended record low level, (b) field *B* was almost constant at a low value, (c) the solar wind velocity reached at its lowest, (d) the tilt angle decreased in the initial two rotations in continuation to the preceding trend, but it increased very rapidly during the next three solar rotations.

- We suggest that, in addition to larger values of diffusion coefficients and drift velocity, the reduced solar wind convection is a significant contribution responsible for the record high GCR intensity during the recent minimum of cycle 23.

**Acknowledgement** We acknowledge the use of solar and plasma/field data available through the NASA/**GSFC** OMNI Web interface. We thank Ilya Usoskin for the Oulu Neutron monitor data and Todd Hoeksema for the HCS inclination data. The authors thank the referee, whose comments and suggestions helped us to improve the paper.